\documentclass[aps,preprint,amsfonts,showpacs,superscriptaddress]{revtex4} 
\pdfoutput=1

\pdfoutput=1
\usepackage{bm}
\usepackage{epsfig,amsopn}
\usepackage{graphicx}
\usepackage{amssymb}
\usepackage{color}
\def\ql{ {Q_{ l}} }
\def\qr{ {Q_{ r}} }

\newcommand \blue{\color{blue}}

\newcommand \bea{\begin{eqnarray}}
\newcommand \eea{\end{eqnarray}}
\newcommand \f{\frac}
\newcommand \nn{\nonumber}
\newcommand \ra{\rangle}
 \newcommand \la{\langle}
\newcommand \p{\partial}

\newcommand\bk{{\bf k}}

\begin{document}

\title[Anomalous transport and current fluctuations in a model of diffusing Levy walkers]
{Anomalous transport and current fluctuations in a model of diffusing Levy walkers}

\author{Abhishek Dhar$^1$ and Keiji Saito$^2$}
\affiliation{$^1$International  Centre for Theoretical Sciences, TIFR, Bangalore 560012, India}
\affiliation{$^2$Department of Physics, Keio University, Yokohama 223-8522, Japan}

\begin{abstract}
A Levy walk is a non-Markovian stochastic process in which the elementary steps of the walker consist of  motion with constant speed in randomly chosen directions and for a  random period of time. The time of flight is chosen from a long-tailed 
distribution with a finite mean but an infinite variance. Here we consider an open system with boundary injection and removal of particles, at prescribed rates, and study the steady state properties of the system. In particular, we compute density profiles, current and current fluctuations in this system. We also consider the case of a finite density of Levy walkers on     
the ring geometry. Here  we   introduce a size dependent cut-off in the time 
of flight distribution  and consider properties of current fluctuations

\end{abstract}

\maketitle

\section{Introduction}
The Levy walk model is a well-studied model to describe anomalous diffusion, in particular super-diffusion \cite{shlesinger86,klafter93,zumofen93,metzler99}.
Among experimental systems the Levy walk model has  been used to model the motion of photons in random medium \cite{barthelemy07}
 and  the diffusion of cold atoms \cite{sagi}. The migration of various species of animals has also been proposed to follow Levy walk statistics though  recent observational data has  questioned  this \cite{andrew07}.
A closely related model is that of  Levy flights \cite{shlesinger86} and fractional diffusion. For the Levy flight model, each elementary step takes the same time but the step length is chosen from a long-tailed distribution. Then, by definition, the second moment of the position of the walker diverges at any time. 

  Most studies of the Levy walk so far have focused  on the motion of a single Levy walker  and a description of the propagator which tells us how an initially localized distribution spreads in time. This is known to be non-Gaussian, with a standard distribution $\sigma$ that grows with time $t$ as $\sigma^2 \sim t^\gamma$ with $1<\gamma <2$.  
One can also consider the many-particle situation where we could, for example, be interested in an open system where Levy walkers can enter or leave the system at the boundaries with prescribed rates. For different rates of injection and emission one can have current-carrying steady states.  One could also consider a ring geometry with a finite density of walkers 
and ask questions on relaxation and fluctuations. These questions are much less studied and is the focus of this paper. 

The problem of steady state transport in a system of non-interacting Levy walkers arises naturally in the case of anomalous heat transport in one-dimensional systems {\blue{\cite{LLP03, dhar08}.}}  A number of recent 
studies indicate that  a  good description of anomalous heat conduction in
one dimensional systems is obtained by modeling the motion of the heat carriers
as  Levy random walks instead of simple random walks \cite{denisov03,metzler04,cipriani05,lepri11,dhar13}. 
Some of the indicators of anomalous heat transport include --- (i) in steady states the dependence of the  heat  current $J$ on  system size $L$ shows the scaling behaviour $J \sim L^{-1+\alpha}$ with 
$\alpha > 0$, (ii) the temperature profiles across systems in  
nonequilibrium steady states are found to be nonlinear, even for very small applied temperature differences 
and, (iii) the spreading of heat pulses in anharmonic chains is super-diffusive.
All these features seem to be captured by the Levy walk description \cite{dhar13}. 
The problem of Levy flights in bounded domains has also been studied  in the context of first passage time distributions and hitting probabilities \cite{buldyrev01,zoia07,majumdar10} and these are also related to some of the steady state properties of the Levy walk model studied here . 

In this paper, we  expand on our earlier work in \cite{dhar13} where  several exact results on the Levy walk model were presented in the context of steady state anomalous heat transport. We show that the steady state current has a power law dependence on the system-size and is non-locally connected to  the  density gradient in contrast to   normal diffusive transport. We also derive the exact cumulant generating function of current for both the open and the ring geometry. These exact results are consistent with numerical observations that have earlier been found for mechanical  models for one-dimensional heat conduction. We believe our analysis will be helpful in discussions of  other types of superdiffusive transport such as Levy transport of light  in random medium \cite{barthelemy07}.

The plan of the paper is as follows. In Sec.~(\ref{free}) we first consider a single Levy walker on the infinite line. We define the precise model studied here and discuss propeties of the Levy propagator and also various moments of the distribution. In Sec.~(\ref{steady}) we discuss the setting up of the transport problem in an open system of many non-interacting Levy walkers. We show how steady state properties like density profile and average current in the system can be computed.   In Sec.~(\ref{CGF}) we discuss current fluctuations and show how the cumulant generating function can be computed exactly. Current fluctuations on a ring geometry are discussed in Sec.~(\ref{ring}) and we conclude with a discussion in Sec.~(\ref{summary}).

\section{Levy diffusion on the infinite line} 
\label{free}
We consider first a Levy walk on the infinite line.
Each step of the walk consists in choosing the step length $x$ and the time 
for the step $t$ from the joint distribution $\eta(x,t)$. Thus $\eta(x,t) dx dt$ is the probability that a step has length between $x$ and $x+dx$ and is of a 
duration  between $t$ and $t+dt$. Here we consider the distribution
\bea
\eta(x,t) = \f{1}{2} \left[\delta(x-vt)+\delta(x+vt)\right]~ \phi(t) 
\eea
Our choice of the step distribution corresponds to choosing a time of flight 
from the distribution $\phi(t)$ and then moving at  speed $v$ in either 
direction, with equal probability. We define  
\begin{equation}
\psi(t)=\int_t^\infty d t'~\phi(t') \label{psidef}
\end{equation}
 as the probability of choosing a time of flight $\geq t$ and
\begin{equation}
\chi(t)=\int_t^\infty d\tau~\psi(\tau)~.\label{chidef}
\end{equation}
We would like to find the propagator $P(x,t)$ which gives the probability 
that the walker is between  $x$ and $x+dx$ at time $t$, given that it was at $x=0$ at time $t=0$. 
To this end first 
let us define $Q(x,t)~ dx~ dt$ be the probability that the walker has precisely landed in 
the interval $dx$ during the time interval $dt$. Note that this does not
include trajectories which were crossing $x$ at time $t$. 
\begin{figure}
\includegraphics[scale=0.7]{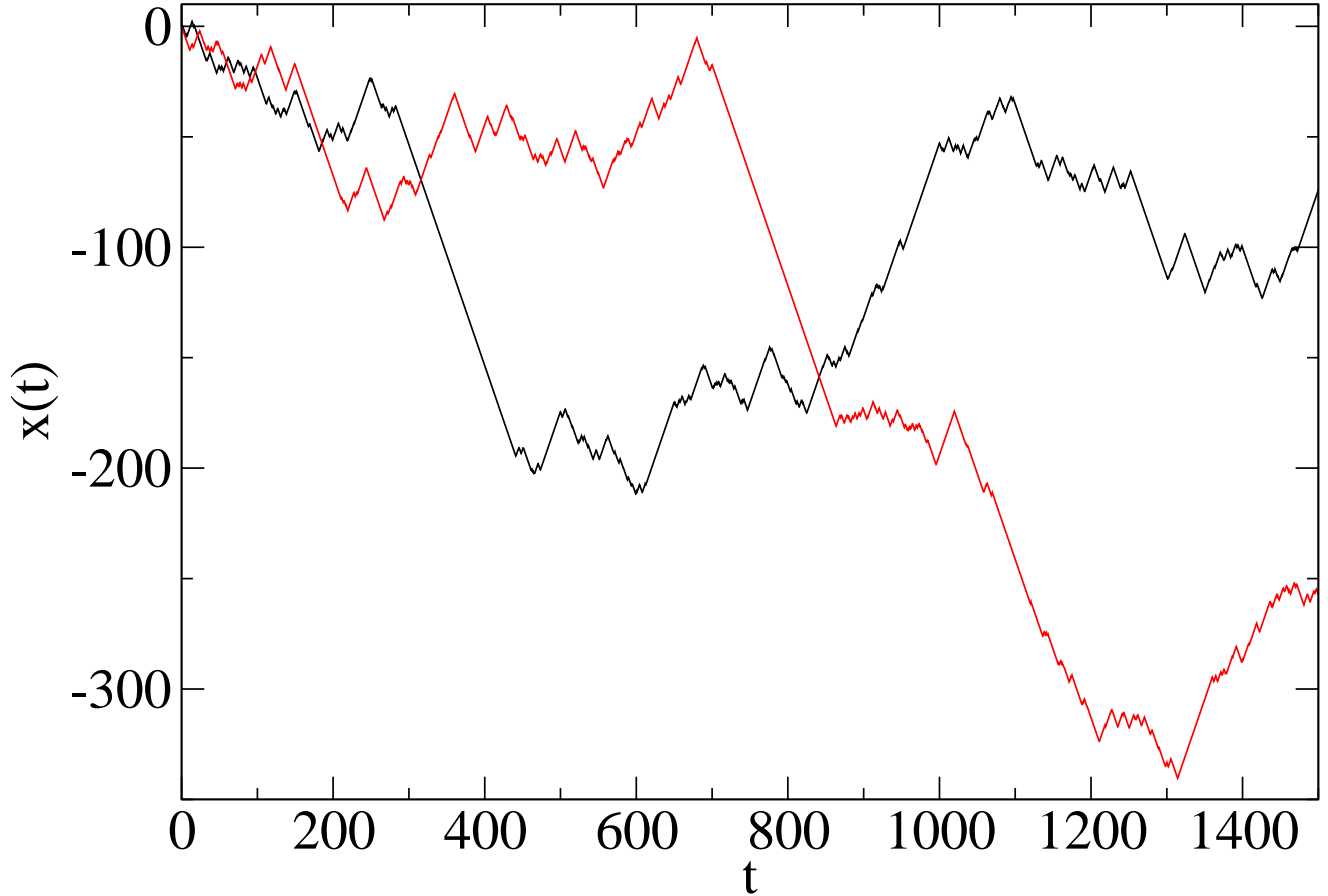}
\caption{Plot of typical trajectories of the Levy walk in one-dimension where the flight time distribution is given by Eq.~(\ref{phiform}) with $\beta=1.5$.}
\label{fig1}
\end{figure}  
We then have the following equations:
\bea
Q(x,t) &=& \int_{-\infty}^\infty dx' \int_0^t ~dt' ~Q(x-x',t-t')~\eta(x',t')+\delta(x)~\delta(t), \label{eq1}\\
P(x,t)&=&\int_{-\infty}^\infty dx' \int_0^t dt' Q(x-x',t-t')~\xi(x',t')~\label{eq2}, \\
{\rm where}~~ \xi(x,t)&=&\f{1}{2}~\left[\delta(x-vt)+\delta(x+vt)\right]~\psi(t)~\label{eq3} 
\eea
is the probability that during a single step 
(starting from the origin), the walker is in the region $x-x+dx$ at time $t$. 
Let us define the Fourier-Laplace transform of a function $f(x,t)$ as:
\bea
\widetilde{f}(k,s)=\int_{-\infty}^\infty dx \int_0^\infty dt~ e^{i k x}~ e^{-st} ~f(x,t) ~.\nn
\eea
Then Eqs.~(\ref{eq1},\ref{eq2},\ref{eq3}) give:
\bea
\widetilde{Q}(k,s)&=&\widetilde{Q}(k,s)~\widetilde{\eta}(k,s)+1 \nn \\
\widetilde{P}(k,s) &=& \widetilde{Q}(k,s)~\widetilde{\xi}(k,s) \nn~, 
\eea
where $\widetilde{\eta}(k,s)= [\widetilde{\phi}(s-ivk)+\widetilde{\phi}(s+ivk)]/2$,~$\widetilde{\xi}(k,s) =[\widetilde{\psi}(s-ivk)+\widetilde{\psi}(s+ivk)]/2$.
Hence we get
\bea
\widetilde{P}(k,s)=\f{\widetilde{\xi}(k,s)}{1-\widetilde{\eta}(k,s)} 
=\f{\widetilde{\psi}(s-ikv)+\widetilde{\psi}(s+i k v)}{2-\widetilde{\phi}(s-i k v)-\widetilde{\phi}(s+i k v)}~,
\eea 
$\widetilde{\psi}(s) =\int_0^\infty dt e^{-st} \psi(t) =[1-\widetilde{\phi}(s)]/s$.
This result also directly follows from noting that the probability  $P(x,t)$ 
satisfies
\bea
P(x,t)= {1 \over 2}\psi(t) \delta(|x|- v t )  
 + {1 \over 2} \int_0^t   d t' \phi(t')  [P(x- v t',t-t') + P(x+ v t',t-t') ]~.
\nn
\eea
Here we   consider Levy walkers with a time-of-flight distribution 
\bea
~\phi(t) &=& \f{\beta}{t_o}\f{1}{(1+t/t_o)^{\beta+1}}~,~~1<\beta<2~. \label{phiform}
\eea
which decays like    a power law   
$ \phi(t) \simeq A ~t^{-\beta-1} $ with $A=\beta t_0^\beta$ at large times. 
 For  this  range of $\beta$  the mean flight time 
 $\la t \ra =\int_0^\infty dt ~t ~ \phi(t)= t_0/(\beta-1) $ is finite but   $\langle t^2 \rangle = \infty$.

\begin{figure}
\includegraphics[scale=0.7]{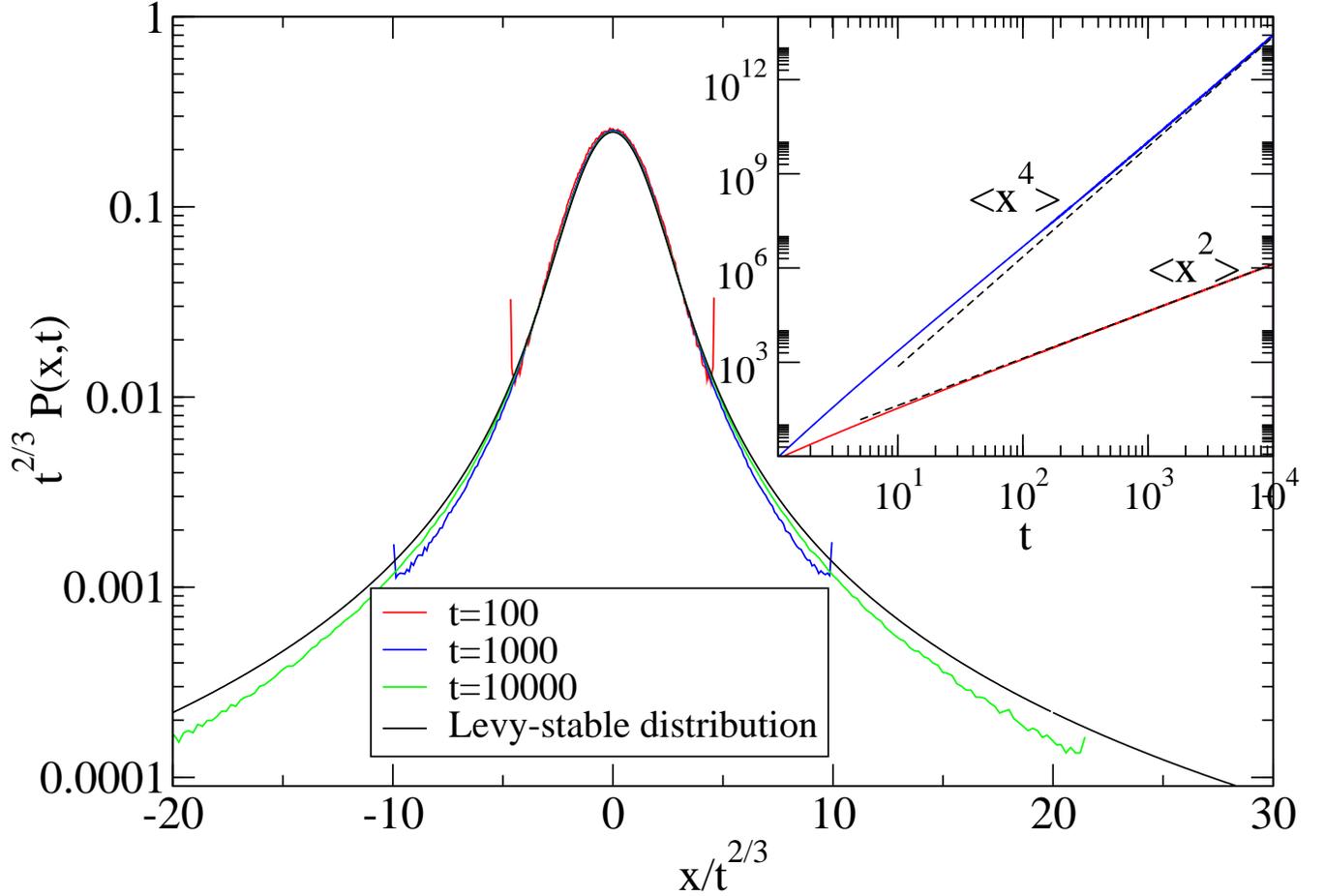}
\caption{Plot of the scaled distribution $t^{2/3}P(x,t)$ versus $x/t^{2/3}$ of the Levy walk on the open line for $\beta = 3/2$ at three different times. Also shown is a plot of the Levy-stable distribution (see text). The inset shows a plot of the mean square displacement and the fourth moment and a comparision with the exact asympotic forms (dashed lines) given by Eqs.(\ref{xsq},\ref{xqd}).  In all plots the time $t_o$ and $v$ are set to one. }
\label{fig2}
\end{figure}

For asympotic properties it is useful to find the form of $\widetilde{P}(k,s)$ for small $k,s$. The laplace transform $\tilde{\phi}$ is given by:
\bea
\tilde{\phi}(s) &=&\int_0^\infty dt~e^{-st}~\phi(t)=1-\la t \ra ~s+
b~\beta (s t_o)^\beta+\cdots ~,  \\
{\rm where}~~b &=& \f{1}{\beta (\beta-1)}~\int_0^\infty~ dz~e^{-z} z^{1-\beta}= \f{1}{\beta(\beta-1)} \Gamma(2-\beta)~. \nn 
\eea
 Hence we get:
\bea
\widetilde{P}(k,s)= \f{1  - c [(s-ikv)^{\beta-1} +(s+ikv)^{\beta-1}]}
{s  - c [(s-ikv)^{\beta} +(s+ikv)^{\beta}]}~, \label{propks}
\eea
where $c=b A/(2 \la t\ra)$.
Taking the inverse Fourier-Laplace transform of this gives us the propagator 
of the Levy walk on the infinite line.  This   
 corresponds to a pulse whose central region is a Levy-stable distribution with a scaling $ x \sim t^{1/\beta}$. This can be seen by expanding Eq.~(\ref{propks}) for $vk/s <<1$ to get $\widetilde{P}(k,s)=[s-c \cos (\beta \pi/2) (vk)^\beta]^{-1}$.  The difference with the Levy-stable distribution is that the Levy-walk propagator has ballistic peaks  of  magnitude $t^{1-\beta}$ at $x=\pm vt$  and vanishes outside this.  The overall behaviour of the propagator is as follows \cite{klafter93}: 
\bea
P(x,t) &\sim t^{-1/\beta} ~e^{{-a x^2}/{t^{2/\beta}}} ~~~~|x| \lesssim t^{1/\beta} \nn \\
&\sim t~x^{-\beta-1}~~~~~~~~~~ t^{1/\beta} \lesssim |x| < vt \nn \\
&\sim t^{1-\beta} ~~~~~~~~~~~~~~~~|x|= v t \nn \\
&= 0~~~~~~~~~~~~~~~~~~~~ |x| > vt ~.
\eea
The time evolution of the Levy-walk propagator, obtained from direct simulations of the Levy walk, is shown in 
Fig.~(\ref{fig2}). We also plot the Levy-stable distribution obtained by taking the Fourier transform of $P(k,t)=e^{-c \cos (\beta \pi/2) |k|^{\beta}t}$.

Various  moments of the 
distribution can be found since 
\bea
\la x^n \ra(t)=  \f{d^n}{d (ik)^n} \int dk e^{ikx} P(x,t) \Big|_{k=0}~,  \nn
\eea
while its Laplace transform is given by
\bea
\la x^n \ra(s) &=& \f{d^n}{d (ik)^n} ~ \widetilde{P}(k,s)~\Big|_{k=0}~. \nn
\eea
Using Eq.~(\ref{propks}) we get the following leading behaviour
\bea
\la x^2 \ra (s) &\simeq& 4 (\beta-1) c v^2 s^{\beta-4}~,\nn \\
\la x^4 \ra (s) &\simeq& 8 (\beta-1) (2-\beta) (3-\beta) c v^4  s^{\beta-6}~, \nn
\eea 
and this implies (for large $t$) \cite{dhar13}
\bea
\la x^2 \ra  &\simeq &   {2~ A ~ v^2\over (3-\beta)(2-\beta) \beta   ~ \la \tau \ra} \,  t^\gamma,~~~~\gamma={3-\beta}~, \label{xsq}   \\ 
\la x^4 \ra  
&  \simeq &   
{  
{4~ A ~ v^4 \over (5-\beta)(4-\beta) \beta   ~ \la \tau \ra} \,  t^{\gamma+2} ~.
} 
 \label{xqd}
\eea 
 We see that for $1<\beta<2$  the motion is  superdiffusive  \cite{zumofen93,metzler99}. 

{Higher dimensions}:
Our results for the Levy walk model are easy to generalize to higher dimensions.Let us consider a walk where in each step the walker chooses a flight time from the same distribution $\phi(t)$ while the joint distribution $\eta (\bar{x},t)$ is given by
\bea
\eta({\bf{x}},t)= \f{\delta(|\bf{x}|-vt)}{ [2 (\pi)^{d/2}/ \Gamma(d/2)]~|{\bf x}|} ~\phi(t)
\eea
 Corresponding to the one-dimensional equations Eqs.~(\ref{eq1},\ref{eq2}) for a Levy walk on the open line, we now have the following equations:
\bea
Q({\bf x},t) &=& \int_{-\infty}^\infty d{\bf x}' \int_0^t ~dt' ~Q({\bf x}-{\bf x}',t-t')~\eta({\bf x}',t')+\delta({\bf x})~\delta(t), \label{eq4}\\
P({\bf x},t)&=&\int_{-\infty}^\infty d{\bf x}' \int_0^t dt' Q({\bf x}-{\bf x}',t-t')~\xi({\bf x}',t')~\label{eq5}, \\
{\rm where}~~ \xi({\bf x},t)&=&\f{1}{2}~\f{\delta(|{\bf x}|-vt)}{[(2 \pi)^{d/2}/ \Gamma(d/2)]~ |{\bf x}|^{d-1}} ~\psi(t)~.\label{eq6} 
\eea
Taking the Fourier Laplace transform then gives
\bea
\widetilde{P}(\bk,s)=\f{\widetilde{\xi}(\bk,s)}{1-\widetilde{\eta}(\bk,s)} ~,
\eea 
with 
\bea
\widetilde{\eta}(\bk,s) &=& \la \widetilde{\phi}(s-iv |\bk| \cos\theta )\ra=  
\f{\int_0^{\pi} d \theta \sin^{d-2}\theta ~\widetilde{\phi}(s-iv |\bk| 
\cos\theta )}{\int_0^{\pi} d \theta \sin^{d-2}\theta} ~,\nn\\ 
~~~\widetilde{\xi}(\bk,s) &=& \la \widetilde{\psi}(s-iv|\bk|\cos\theta) \ra~,
\eea
and where $\la ...\ra$ denotes an average over the polar angle $\theta$.
Proceeding as for the one-dimensional case we get the analogue of Eq.~(\ref{propks}):
\bea
\widetilde{P}(\bk,s)=\f{1-2c\la( s+i |\bk| v \cos\theta)^{\beta-1}\ra}{  s-2c\la( s+i |\bk| v \cos\theta)^{\beta}\ra}~.
\eea

\section{Levy diffusion in a finite system connected to infinite reservoirs} 
\label{steady}
To study an open system of Levy walkers we consider a finite system connected to two semi-infinite reservoirs on which the density of walkers is maintained at fixed values. 
Thus we consider our system to be the finite segment between $(0,L)$ and this is connected on the two sides to reservoirs.
The left reservoir consists of the region $  x \le 0$ while the right 
reservoir consists of the region $x \geq L $. We set $Q(x,t)=Q_l$ for 
points on the left reservoir and $Q(x,t)=Q_r$ for those on the right. 
In general if we know the distributions $Q(x,\tau)$ and $P(x,\tau)$ for all times 
$-\infty< \tau <t$ then the distribution at time $t$ is given by:
\bea
Q(x,t) =  \int_{-\infty}^\infty  dy \f{1}{2v} ~ Q(y,t-|x-y|/v) ~\phi(|x-y|/v)~, \label{Qeq} \\
P(x,t) = \int_{-\infty}^\infty  dy\f{1}{2 v}~ Q(y,t-|x-y|/v) ~\psi(|x-y|/v)~. \label{Peq}
\eea
In the above expressions $Q(x,t)$ gets contributions from walkers starting from
all possible points $y$ and landing precisely at $x$ at time $t$.  On the other hand $P(x,t)$ gets contributions  from walkers starting at $y$ and being either at or  passing $x$ at time $t$.
Since the distribution $Q(x,t)$ is constrained to take either of the values $Q_l$
or $Q_r$ in the reservoirs, the above equation gives, 
for points on the system
\begin{eqnarray}
Q(x,t) & = & 
 \f{Q_l}{2}~ \psi(x/v) 
+\f{Q_r}{2} ~\psi[(L-x)/v] \nn \\
&+& \int_{0}^L  dy \f{1}{2v} ~ Q(y,t-|x-y|/v) ~\phi(|x-y|/v)~, \label{Qeffeq} \\
P(x,t) &= &\f{Q_l}{2} ~ \chi(x/v) 
+\f{Q_r}{2} ~ \chi[(L-x)/v] \nn \\
&+& \int_{0}^L  dy\f{1}{2 v}~ Q(y,t-|x-y|/v) ~
\psi(|x-y|/v)~, \label{Peffeq}
\end{eqnarray}
where we used the definitions of $\psi$ and $\chi$ from Eqs.~(\ref{psidef}) and (\ref{chidef}). We see that the equations above can be interpreted by thinking of a system where particles  enter from the left boundary with flight times $t$ at a rate $Q_{\ell}~\psi(t)$ and from the right boundary at a rate $Q_r~ \psi(t)$. 

\subsection{Steady state density profile}
\label{ssdens}
In the steady state we have $Q(x,t)=Q(x)$ and $P(x,t)=P(x)$, hence we get:
\begin{eqnarray}
 Q(x)&=&\int_0^L  dy~\f{1}{2v} \phi(|x-y|/v) ~Q(y) + \f{Q_l}{2} \psi(x/v)
+ \f{Q_r}{2} \psi[(L-x)/v]~,  \label{QeqSS} \\ 
 P(x) &=& \int_0^L  dy \f{1}{2v} \psi(|x-y|/v)~ Q(y)  
+  \f{Q_l}{2} \chi(x/v) + \f{Q_r}{2} \chi[(L-x)/v] ~. \label{PeqSS}
\end{eqnarray}
It can be verified by direct substitution that the solution of Eq.~(\ref{QeqSS}) is given by 
\begin{equation}
Q(x)=(Q_l-Q_r) H(x)+ Q_r \label{Qsol}
\end{equation}
where $H(x)$ is the probability that a Levy walker starting at position $x$ 
will first hit the left reservoir  (i.e. the region $x < 0$) before it hits the right reservoir (i.e. $x >L$), and 
satisfies 
\begin{equation}
H(x)-\int_0^L  dy~\f{1}{2v} \phi(|x-y|/v) ~H(y) =  \f{1}{2} \psi(x/v)~.
 \label{Heq} 
\end{equation}
If one considers a Levy flight with distribution   $\rho(z)=[\phi(z/v)+\phi(-z/v)]/(2 v)$  of steps $z$, the probability $H(x)$   that  starting at $x$, the  flight hits first the left 
bath satisfies exactly Eq.~(\ref{Heq}).  Hence
 by following the same mathematical steps as in     \cite{buldyrev01} to study equations such as (\ref{Heq}), one  can show  that, in the large $L$ limit,  the solution  $H(x)$   satisfies
\bea
\int_0^L dy~\psi (|x-y|/v)~ {\rm Sgn}(x-y)~ H'(y)= 0~,  \label{Qpeq}
\eea
with $H(0)=1$ and $H(L)=0$. For  a $\phi(\tau)$    decaying as in (\ref{phiform}), the solution of Eq.~(\ref{Qpeq})  is \cite{dhar13} 
\bea 
H'(x)=-B [x (L-x)]^{\beta/2-1}~. \label{Qpsol}
\eea
Integrating this and imposing the boundary conditions $H(0)=1$ and 
$H(L)=0$,  one obtains the constant  $B= 
\Gamma(\beta)/  
\Gamma(\beta/2)^2~
L^{1- \beta}$. From Eq.~(\ref{Qsol}) we then finally get $Q(x)$.  
Substituting $\psi(x/v)=-vd\chi(x/v)/dx$ in (\ref{PeqSS}) we obtain $P(x)=\chi(0)Q(x)-\int_0^L dx' \chi(|x-x'|/v) {\rm Sgn} (x-x') Q'(x')/2$. Using Eq.~(\ref{Qpsol}) we then get in the limit of large L:
\bea
P(x)=\chi(0)Q(x) =  \langle \tau \rangle Q(x) ~. \label{pq}
\eea 
In Fig.~(\ref{fig1}), we  compare  numerical results obtained by solving Eqs.~(\ref{QeqSS},\ref{PeqSS}) 
 with the exact results of  Eqs.~(\ref{Qpsol},\ref{pq}). The  profiles are nonlinear and  
 look   similar to those observed for temperature profiles in 1D heat conduction \cite{LLP03,dhar08}. 
\begin{figure}
\includegraphics[scale=1.6]{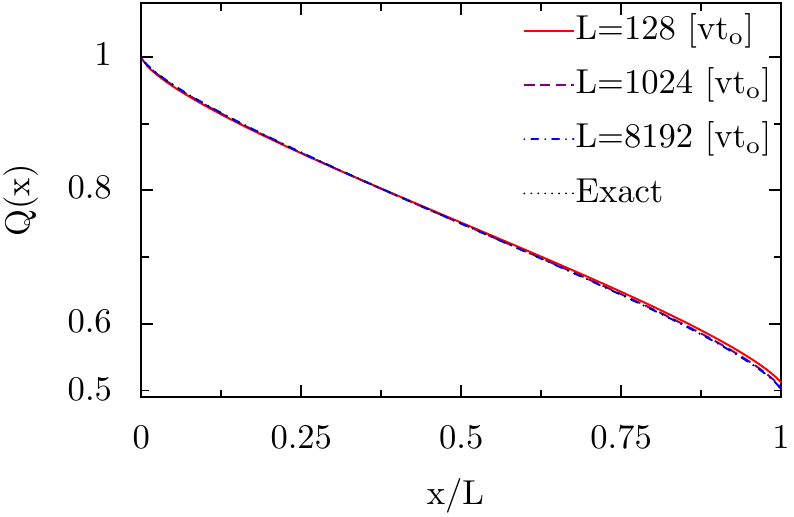}
\caption{Plot of $Q(x)$  for different system sizes, for the Levy walk with $\beta = 1.5$, $\ql=1.0, \qr= 0.5$. The data are obtained by solving Eqs.~(\ref{QeqSS},\ref{PeqSS}) with discretized space.  The distribution of flight times is $\phi (\tau)=\beta/t_o/(1+\tau/t_o)^{\beta +1}$.
}
\label{fig1}
\end{figure} 

\subsection{Steady state current}
\label{sscurr}
To derive an expression for the current operator, we need to write a continuity equation of the form  $\p P(x,t)/\p t+\p J(x,t)/\p x=0$. To this end let us make a change of the integration variables in Eqs.~(\ref{Qeq},\ref{Peq}) from $y$ to $t'=t-|x-y|/v$ to get the following forms:
\bea
Q(x,t)=\f{1}{2} \int_{-\infty}^t d t' [~Q(x-vt+v t',t') +Q(x+vt-vt',t')~]~\phi(t-t')~,\label{Qeq2} \\
P(x,t)=\f{1}{2} \int_{-\infty}^t d t' [~Q(x-vt+v t',t') +Q(x+vt-vt',t')~]~\psi(t-t')~.\label{Peq2}
\eea
 Taking a time-derivative of Eq.~(\ref{Peq2}) we get
\bea
&\f{\p P(x,t)}{\p t}= Q(x,t) \psi(0) \nn \\ 
 &~~~+ \f{1}{2} \int_{-\infty}^t d t'  \left( -v\f{\p}{\p x} \right) \left[~Q(x-vt+v t',t') -Q(x+vt-vt',t')~\right]~\psi(t-t') \nn \\
&~~~+ \f{1}{2} \int_{-\infty}^t d t' [~Q(x-vt+v t',t') +Q(x+vt-vt',t')~]~\f{d}{d t}\psi(t-t')~. \nn
\eea
Using $\psi(0)=1$, $-d\psi/dt=\phi$ and Eq.~(\ref{Qeq2}), the first and last terms on the right hand side vanish, and we get the current continuity equation with
the following form of the current operator 
\bea
 J(x,t) = \f{1}{2}\int_{-\infty}^{\infty} dy  ~ Q(x-y,t-|y|/v)  ~{\rm Sgn} (y)~ \psi(|y|/v) ~. \label{currexp}
\eea
This equation is easy to understand physically. 
The contribution to the integral coming from $y >0$ corresponds to 
particles crossing the point $x$ from left to right  which started their flight at  $x-y$ at time $t-y/v$ (the factor  $\psi(y/v)$  comes from the fact that these particles  have a flight time longer than $y/v$). Similarly 
the other part of the integral (from $y<0$) corresponds to a right-to-left 
current. 

In the steady state, setting $Q(x,t)\equiv Q(x)$  we get the result 
\bea
 J(x) = \f{1}{2}\int_{-\infty}^{\infty} dy  ~ Q(x-y)  ~{\rm Sgn} (y)~ \psi(|y|/v) ~. \label{sscurr}
\eea
Using the values of $Q$ in the reservoirs and the 
steady state solution given by Eqs.~(\ref{Qsol},\ref{Heq}) we evaluate the 
curent at $x=0$ and obtain
\bea
J=\f{(Q_l-Q_r)}{2}~\left[ ~\int_0^\infty dy ~\psi(y/v) - \int_0^L dy ~H(y) \psi(y/v)~\right]~. \label{jeq}
\eea

Since the system is non-interacting, this result can be obtained directly by 
noting   that the current is due to particles which  enter from the left
and leave to the right ( and to  the symmetric contribution). 
We then simply need to know the rate at which the non-interacting particles 
enter the system on the left side and leave the system into the right 
reservoir. This is given by
\bea
p_l&=&\f{Q_l}{2}\int_{-\infty}^0 dy~\int_{(L-y)/v}^\infty d \tau \phi(\tau) \nn \\
&+& 
\f{Q_l}{2} \int_{-\infty}^0 dy~ \int_{0}^{L} \f{dx}{v} ~[1-H(x)]~\phi[(x-y)/v]~, 
\eea
with a similar  expression for the right to left rate $p_r$. The net current given by $J=p_l-p_r$ is easily seen to be identical to Eq.~(\ref{jeq}).

After a 
partial integration  and using the fact that
$Q(0) = \ql$ and $Q(L) = \qr$, one gets \cite{dhar13}
\bea
J(x)&=& -\f{v}{2} \int_{0}^{L} dy~ \chi (|x-y| / v)~ Q'(y)~. \label{jss}
\eea
We note that $dJ/dx=0$ gives Eq.~(\ref{Qpeq}) and so the current is independent of $x$, as  expected.   
Evaluating the current at $x=0$ and  using  
Eq.~(\ref{Qpsol}), we get for large $L$ \cite{dhar13}
\bea
J \simeq   ( \ql- \qr)  \  
{ A~ v^{\beta} 
~\Gamma(\beta )~ \Gamma(1- {\beta \over 2}) \over 2 ~\beta (\beta-1)~\Gamma({\beta \over 2}) 
}   \  L^{\alpha -1}  ,~~~\alpha= 2-\beta~ .
 \label{Jsol}
\eea
Thus unlike in normal diffusive systems where $J \sim 1/L$ here we have a anomalous 
conductivity exponent $\alpha \neq 0$.
From Eq.~(\ref{xsq}) we then get the relation $\alpha=\gamma-1$, 
 between the conductivity exponent 
for transport and the exponent for 
Levy-walk diffusion. In the context of anomalous heat conduction this relation, between the exponents for the Levy model, was noted in 
 \cite{denisov03}, numerically observed in $1D$ heat conduction models \cite{cipriani05,zhao} and a derivation based on linear response theory has recently been proposed  \cite{liuli}.

In the  large $L$ limit by using Eq.~(\ref{pq}) in Eq.~(\ref{jss}) 
we obtain \cite{dhar13}
\bea
J&=&-\f{v}{2 \la  \tau \ra }\int_0^L d{ y}~ \chi(|x - y| /v) P'(y )~. 
\eea
This is the analogue of the usual diffusion equation $J=-D\p P(x)/\p x$ in the case of diffusive systems 
and  can be interpreted as  current being  non-locally  connected to the density gradient.

\section{Current fluctuations in the open system}
\label{CGF} 
Since the particles are independent the current fluctuations can be described by a   Poissonian process
characterized by the rate at which 
walkers injected  from the left reservoir   end up  
(either after a  non stop flight or  a non direct flight) into the right reservoir or walkers injected from the right reservoir end up   into the left reservoir.
In the steady state let us  look at the cumulant generating function
\bea
Z(\lambda)=\la e^{\lambda Q} \ra~, 
\eea
where $Q$ is the net total number of particles that have crossed from the left reservoir  into the system in the time interval $0-\tau$. The total current is carried by independent Levy walkers and we can identify contributions from three distinct independent processes, namely ---

(1)  A particle entered the system at some $s<0$  from the left  reservoir with a flight time $t'<L/v$ and then   exits into the 
left reservoir at some time $t$ in the  interval $0-\tau$.
 We define a random variable $z_\ell(t',s,\tau)$ which has value $1$ 
whenever this process occurs and is otherwise $0$. Similarly $z_r(t',s,\tau)$ is defined for the process when the particle enters from right reservoir at $s<0$ and exits into left reservoir at $t \in (0,\tau)$.  

(2) A particle enters the system from the left reservoir at some time $s \in (0,\tau)$   with a flight time $t'$ and thereafter does not exit back into the left reservoir. We define a variable $y_\ell(t',s,\tau)$ which has value $1$ whenever this process occurs and is otherwise zero.  

(3) A particle enters the system from the right reservoir at time $s \in(0,\tau)$ with a flight time $t'$   and  then exits into the left reservoir some time during the remaining time  interval. We define a variable $y_r(t',s,\tau)$ which has value $1$ whenever this process occurs and is otherwise zero.  

Let us divide the time interval $(0,\tau)$ into small intervals of size  $ds$.
Hence we can write $Q=\sum_{s=-\infty}^0 ~ \sum_{t'=0}^{L/v} [~z_\ell(t',s,\tau) + z_r(t',s,\tau)~] + \sum_{s=0}^\tau \sum_{t'=0}^\infty [~ y_\ell(t',s,\tau)+ y_r(t',s,\tau)~]$.
Since the particles are all independent we can write
\bea
Z(\lambda)&=&  \prod_{s=-\infty}^0~\prod_{t'=0}^{L/v} 
\left\la e^{-\lambda  z_\ell(t',s,t)} \right\ra ~\left\la e^{-\lambda  z_r(t',s,t)} \right\ra ~   \nn \\
&\times&\prod_{s=0}^\tau~ \prod_{t'=0}^\infty \left\la e^{\lambda y_\ell(t',s,\tau)} \right\ra  \left\la e^{-\lambda y_r(t',s,\tau)} \right\ra~.
\eea

Let us define $\alpha_\ell(t),\alpha_r(t)$ to be the rates at which particles enter from the left and right reservoirs respectively with flight times $t$.  Let $\gamma_{\ell,\ell}(t',t)$ be the rate of escape  into the left reservoir at time $t$,  of particles that entered the system from the left reservoir at time $t=0$ with a time of flight $t'$, while 
$\gamma_{r,\ell} (t',t)$ is the rate of escape into the left reservoir at time $t$ of a particle which entered the system from the right reservoir at time $t=0$ with a flight time $t'$. Similarly we define $\gamma_{r,r}(t',t)$ and $\gamma_{\ell,r}(t',t)$. 

Then we have
\begin{eqnarray}
\prod_{s=-\infty}^0~\prod_{t'=0}^{L/v} 
\la e^{-\lambda  z_\ell(t',s,\tau)} \ra  
&=&   \prod_{s=-\infty}^0 ~\prod_{t'=0}^{L/v} \{ e^{-\lambda} \alpha_\ell(t') ~ds dt'~ D_{\ell,\ell}(t',s,\tau)  
\nn \\
&&+ [1- \alpha_\ell(t') ~ds dt' D_{\ell,\ell}(t',s, \tau)~] \} \nn \\ 
&=& 
e^{(e^{-\lambda}-1)~\int_{s=-\infty}^0 ds~~ \int_0^{L/v}  dt'~ \alpha_\ell(t') ~D_{\ell,\ell}(t',s,\tau)~}, \label{exp1}  \\
\prod_{s=-\infty}^0~\prod_{t'=0}^{L/v} 
\la e^{-\lambda  z_r(t',s,\tau)} \ra  
&=&   \prod_{s=-\infty}^0 ~\prod_{t'=0}^{L/v} \{ e^{-\lambda} \alpha_r(t') ~ds dt'~ D_{r,\ell}(t',s,\tau)  
\nn \\
&&+ [1- \alpha_r(t') ~ds dt' D_{r,\ell}(t',s, \tau)~] \} \nn \\ 
&=& e^{(e^{-\lambda}-1)~\int_{s=-\infty}^0 ds~~ \int_0^{L/v}  dt'~ \alpha_r(t') ~D_{r,\ell}(t',s,\tau)~}, \label{exp2}
\end{eqnarray}
where $D_{\ell,\ell}(t',s,\tau)$ ($D_{r,l}$) is the probability that a particle which entered the system from the left (right) reservoir at time $s$ with a flight time $t'$ is emitted  into the left reservoir during the time interval $(0,\tau)$. This is given by
$D_{\ell,\ell}(t',s,\tau)=\int_0^{\tau} dt S_\ell(t',t-s) \gamma_{\ell,\ell}(t',t-s)~$ where 
$S_\ell(t',t)=e^{-\int_0^{t} ds (\gamma_{\ell,\ell}(t',s)+\gamma_{\ell,r}(t',s))}$ is the probability of survival upto time $t$ of a particle entering the system from the left reservoir at time $t=0$ with a flight time $t'$. 

Similarly we have 
\begin{eqnarray}
\prod_{s=0}^{\tau}  \prod_{t'=0}^{\infty} \left\la e^{\lambda y_\ell(t',s,\tau)} \right\ra  
&=& \prod_{s=0}^{\tau} ~\prod_{t'=0}^{\infty} ~e^{\lambda} ~\alpha_\ell(t') ds dt' ~S_{\ell,\ell}(t',\tau-s) \nn
\\&& + [~1- \alpha_\ell(t') ds ~dt'~S_{\ell,\ell}(t',\tau-s)~] \nn \\
&=&e^{ (e^{\lambda}-1)~\int_{0}^\tau ds \int_0^{\infty} dt'~\alpha_\ell(t')   
~S_{\ell,\ell} (t',\tau-s)}~, \label{exp3} \\
\prod_{s=0}^{\tau} \prod_{t'=0}^{\infty} \left\la e^{-\lambda y_r(t',s,\tau)} \right\ra
&=& \prod_{s=0}^{\tau} ~\prod_{t'=0}^{\infty}~e^{-\lambda}~  \alpha_r(t') ds dt'~D_{r,\ell}(t',\tau-s) \nn \\
&&+ [~1- \alpha_r(t') ds ~D_{r,\ell}(t',\tau-s)~]\nn \\
&=&e^{ (e^{-\lambda}-1)~\int_{0}^\tau ds \int_0^{\infty} dt'~ \alpha_r(t')  ~D_{r,\ell}(t',\tau-t)}, \label{exp4}
\end{eqnarray}
 where $S_{\ell,\ell}(t',s)$ is the probability that a particle entering the system from the left reservoir at time $s=0$ with a flight time $t'$ 
does not exit into the left reservoir in the time interval $0-s$ and $D_{r,\ell}(t',s)$ is the probability that a particle  entering the system from the right reservoir at time $s=0$, with a flight time $t'$, exits into the left reservoir in the interval $0-s$. 

We now note that in Eq.~(\ref{exp1}),  $\int_0^{L/v}  dt'~ \alpha_\ell(t') ~D_{\ell,\ell}(t',s,\tau)$ is the average rate at which particles injectd into system from left reservoir at time $s$ are injected back into the left reservoir between times $0-\tau$. Clearly for large $\tau$ large compared to the residence time of a walker inside the system, this will be $O(\tau^0)$ and decay as $e^s$ for large $s$. Hence we expect, that for large $\tau$,
$\int_{s=-\infty}^{0} ds \int_0^{L/v} dt'~ \gamma_\ell(t')~D_{\ell,\ell}(t',s,\tau) \sim O(\tau^0) $ and similarly $  \int_{s=-\infty}^{0} ds \int_0^{L/v} dt' \gamma_r(t')~D_{r,\ell}(t',s,\tau)\sim O(\tau^0) $.
  
On the other hand in Eq.~(\ref{exp3}), $\int_0^{\infty} dt'~\alpha_\ell(t')   ~S_{\ell,\ell} (t',\tau-s)$ is the average rate at which particles injected into system at time $s$ from the left reservoir are not reinjected back into the left reservoir in the time interval $\tau-s$ while, in Eq.~(\ref{exp4}), $\int_0^{\infty} dt'~ \alpha_r(t')  ~D_{r,\ell}(t',\tau-t)$   is the average rate at which particles injected into system at time $s$ from the right reservoir are  injected  into the left reservoir in the time interval $\tau-s$. Both of these quantities are $O(\tau^0)$. Hence for large $\tau$,
\bea
\int_{s=0}^\tau ds \int_0^{\infty} dt'~\alpha_\ell(t')   ~S_{\ell,\ell}(t',\tau-s)=p_L~\tau,~\nn \\
\int_{s=0}^\tau ds \int_0^{\infty} dt'~ \alpha_r(t')  ~D_{r,\ell}(t',\tau-s)=p_R ~\tau,~ \nn 
\eea
where $p_L$ is the rate at which 
walkers injected  from the left reservoir   end up  
(either after a  non stop flight or  a non direct flight) into the right reservoir and $p_R$ is the rate at which  walkers injected from the right reservoir end up   into the left reservoir.

Hence we get  the leading large $\tau$ behaviour as $Z(\lambda) \sim   e^{\mu (\lambda) {\tau} }$ where 
\bea
\mu (\lambda) = p_L (e^\lambda -1) { +} p_R (e^{-\lambda} { -} 1)~. \label{musol}
\eea
In the case $\ql=1$ and $\qr=0$ let ${\cal P}$ be the rate at which walkers,  which will end up into the right reservoir, are  injected from the left reservoir.  For general $\ql$ and $\qr$, because the walkers are independent and because they have no preferred direction, one has 
$ p_ L= \ql {\cal P}$ and $p_R= \qr {\cal P}$. 
Hence we get  for the cumulant generating function of ${\cal Q}$ \cite{dhar13} 
\bea
\mu(\lambda)= J~ \f{\ql (e^\lambda -1) +\qr (e^{-\lambda}-1)}{\ql-\qr} ~,\label{mu}
\eea
{where $J= (\ql-\qr) {\cal P}$.}
We can check that $\mu'(\lambda=0)$ gives the current $J$. 
In Fig.(\ref{fig2}) we  check  the validity of Eq.~(\ref{mu}) from direct simulations of the open system.  From Eq.~(\ref{mu}), for the case $Q_\ell=1,~Q_r=0$ we expect all cumulants $\la Q^n \ra/\tau$ to be independent of $n$ for sufficiently large $\tau$, and we check this for different system sizes.  The simulations are done by injecting particles with flight time $t$ at rates $Q_{\ell}\psi(t)$, $Q_{r}\psi(t)$ from the left and right reservoirs [see discussion after Eqs.~(\ref{Qeffeq},\ref{Peffeq})]. These particles perform Levy walks till they exit from the system.

\begin{figure}
\includegraphics[scale=0.7]{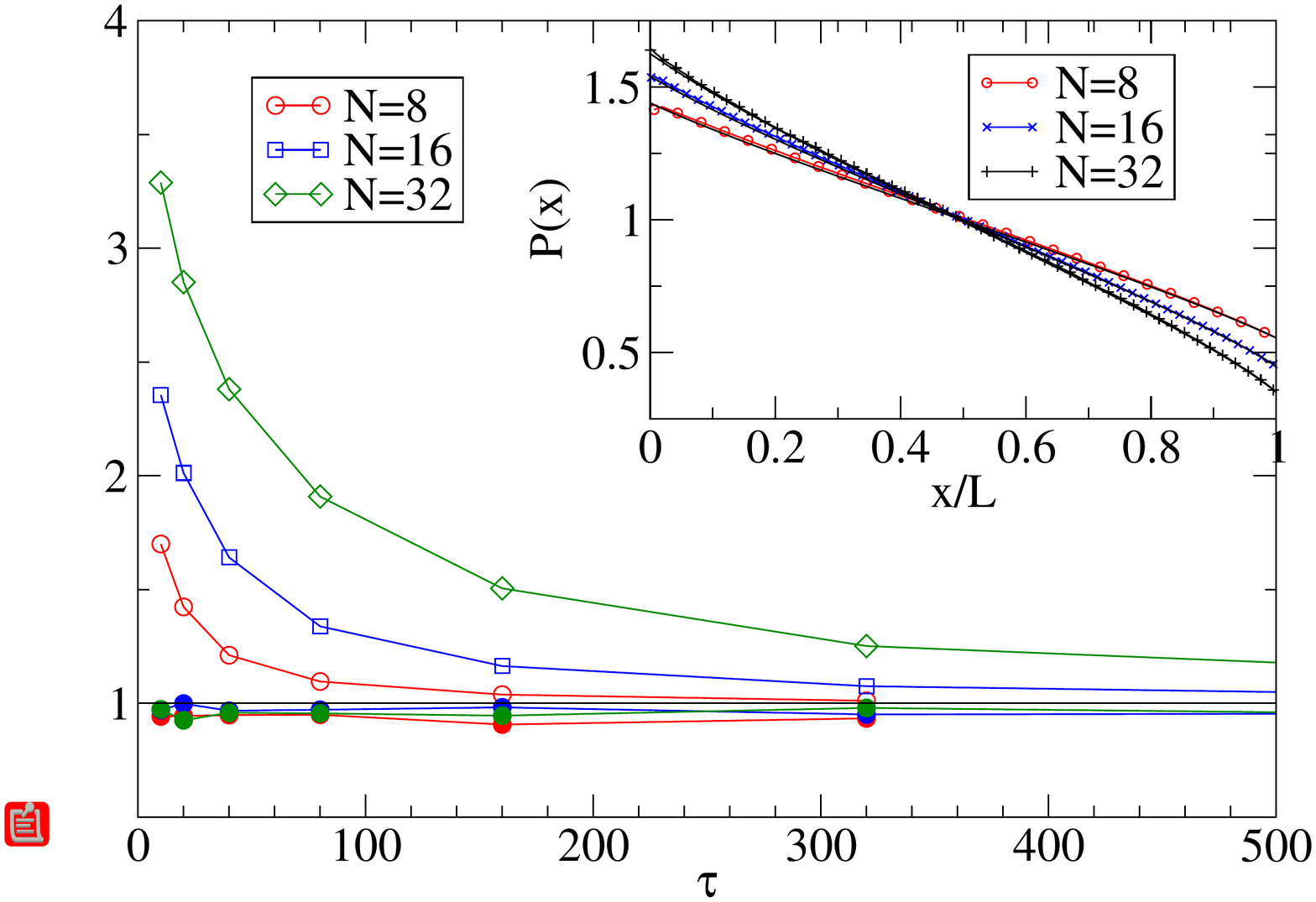}
{ \caption{Monte-Carlo results for $\langle {\cal Q}^2 \rangle_c / 
\langle {\cal Q} \rangle$ and  $\langle {\cal Q}^3 \rangle_c / 
\langle {\cal Q} \rangle$ (filled circles) as a function of measurement time $\tau$
for several system sizes with the parameters $\beta = 1.5$, $(\ql,\qr)=(1,0)$  for the same model as in Figure 1.  
The data 
agree with the results of our theory (\ref{mu}).
The inset shows the density profiles as produced from the Monte-Carlo simulations compared to the exact numerical solutions from Eqs.~(\ref{QeqSS},\ref{PeqSS}) [solid black lines].
}}
\label{fig4}
\end{figure}

Not surprisingly  we also note that the following fluctuation theorem symmetry relation \cite{gc} is satisfied:
\bea
\mu (\lambda)= \mu[-\lambda -(\ln \ql -\ln \qr)]~.
\eea

{\bf Additivity principle:} An interesting observation is that 
the  generating 
function of the integrated current $\mu(\lambda)$ [given by Eq.~(19) in main text] matches exactly 
with the formula obtained from
 the  additivity principle (AP) \cite{BD04}  which 
 gives an expression for $\mu_{AP}(\lambda)$ in terms of the  the conductivity $D$ and equilibrium current 
fluctuations $\sigma$ defined respectively as 
\bea
\begin{array}{l}
D(Q)=\lim_{\Delta Q \to 0} L J  / \Delta Q ~,  \\
\\
\sigma(Q)= L ~{\lim_{t \to \infty} \la {\mathcal {Q}^2}\ra / t} ~.
\end{array}
\eea
The  expression for $\mu(\lambda)$ from AP is
\bea
\mu_{AP}(\lambda) &=& -\f{K}{L}\left[ \int_{\qr}^{\ql} dQ \f{D(Q)}{\sqrt{1+2 K \sigma(Q) }}\right]^2~, \nn \\
{\rm with}~~\lambda &=&\int_{\qr}^{\ql} dQ \f{D(Q)}{\sigma(Q)} 
\left[ \f{1}{\sqrt{1+2 K \sigma(Q) }} -1 \right]~.~~~~\label{muAP}
\eea
(this is a parametric expression: as $K$ varies, $\mu$ and $\lambda$ vary). 
From our exact results for $\mu(\lambda)$ we find $D=L p$ and $\sigma=L~\mu''(\lambda=0)=2 D Q~$. Using these in Eq.~(\ref{muAP}) and after explicitly 
performing the integrals we find $\mu_{AP}(\lambda)= \mu(\lambda)$. This result is somewhat surprising since the additivity principle is expected normally to hold 
for diffusive systems (here $D$ and $ \sigma$ have a $L$-dependence, whereas in usual diffusive systems they don't). 
We note that somewhat similar behavior has been observed in deterministic models where anomalous transort satisfies the AP \cite{sd11}.

\section{Current fluctuations in ring geometry}
\label{ring}
In the ring geometry, the system consists  of  a fixed number $N$  of particles  which perform independent Levy walks  on a ring of length $L$. In the steady state, the density of particles is uniform.
As the walkers are independent, the  cumulants of the integrated current $ {\cal Q} $   are related to those of  the displacement  $x(t) $ of a single   walker on the infinite line (in the steady state). The number of times that a single walker crosses a fixed point in the ring geometry is
approximately $x(t)/L$. Hence in the  steady state where the distribution of particle is uniform, the $n$th order of cumulant is given by
\bea
\la{\cal Q}^n\ra_c  \sim 
{ N \over L^n} \la{ x(t)}^n\ra_c  = {\rho \over  L^{n-1}}  \, , \la{ x(t)}^n\ra_c  \label{Qnxn}
\eea
 where $\rho$ is the density on the ring.
If the walkers perform on the ring the same Levy walks  as on the infinite line, the cumulants of  $ x(t) $ and therefore those of ${\cal Q}$  grow,
as in (\ref{xsq},{\ref{xqd}}), faster than linearly with time 
(same exponent but a different prefactor as, on the ring, the walker is  in  its  steady state  rather than  starting  a flight at $t=0$).

We now introduce the modified ditribution $\phi (t )$ with a cut-off time $t_L \sim L^\delta$. This cutoff can be introduced on physical grounds. For example some possible cut-offs that can be argued are ---

(a) as for the open  geometry, the length of the flights 
cannot exceed the  system size and therefore   $\delta=1$.  

(b) Alternatively  $t_L $ should be of the order of $t^*$, 
the typical relaxation time in the system.  
From the form $P(k,t)=e^{-c \cos (\beta \pi/2) |k|^{\beta}t}$, valid for $vk << 1/t$, 
we can pull out a relaxation time for the shortest wave number on the ring $k=2 \pi/L$, giving $t^* \sim k^{-\beta}$ and this would give $\delta  = \beta$.

(c) Finally from the mean square displacement $\la x^2 \ra \sim t^{3-\beta}$ we obtain a relaxation time $t^*=L^{2/(3-\beta)}$ and this gives $\delta=2/(3-\beta)$. 

With such a  cut-off $\tau_L$, 
the  moments  of the flight times  $ \langle t^n \rangle $ would be finite and the motion would be diffusive. On the infinite line, the propagator can then be expanded as
\bea
\widetilde{P}(k,s)=\f{\sum_{n=1}^\infty (-1)^n \la t^n\ra [(s-ikv)^{n-1}+(s+ikv)^{n-1}]/n!}{\sum_{n=1}^\infty (-1)^n \la t^n\ra [(s-ikv)^{n}+(s+ikv)^{n}]/n!}~. 
\eea
The Laplace transforms of various moments can be computed and are given by
\bea
\la x^2 \ra(s) &\simeq& \f{v^2 \la t^2 \ra}{ \la t \ra} \f{1}{s^2}+ \f{v^2(3 \la t\ra^2 - 4 \la t \ra \la t^3 \ra)}{6 \la t\ra^2} \f{1}{s}, \nn \\
\la x^4\ra (s)&\simeq& \f{6 v^4 \la t^2 \ra^2}{\la t\ra^2} \f{1}{s^3} + \f{v^4(6 \la t^2\ra^3 - 10 \la t \ra \la t^2\ra \la t^3\ra  + \la t\ra ^2 \la t^4\ra)}{  \la t\ra^3} \f{1}{s^2}~. \nn
\eea
One would  then get   for the first two cumulants of $x$ at large $t$ \cite{dhar13}
\bea
 { \la x^2 \ra_{{c}}  \over  v^2 ~t}  \simeq {\la t^2 \ra \over \la t \ra}  
    ~ , ~~~~~  { \la x^4 \ra_c \over v^4 ~ t}  \simeq 
{\la t^4 \ra \over \la t \ra} - 6 { 
 \la t^2 \ra 
 \la t^3 \ra   \over \la t \ra^2} + 3 { \la t^2 \ra^3  \over \la t \ra^3} 
  ~ .
\label{dif-cum}
\ \
\eea
It becomes increasingly cumbersome to evaluate higher moments. For the sixth order cumulant we obtain
\bea
\f{\la x^6 \ra_c}{v^6 t}&=& \f{\la t^6 \ra}{\la t\ra}-15 \f{\la t^2 \ra \la t^5 \ra}{\la t \ra^2}-15 \f{\la t^3 \ra \la t^4 \ra}{\la t \ra^2}+ 60 \f{\la t^2 \ra^2 \la t^4\ra}{\la t \ra^3}+ 90 \f{\la t^2\ra \la t^3\ra^2}{\la t \ra^3}
\nn \\
&-& 150 \f{\la t^2\ra^3 \la t^3\ra}{\la t\ra^4}+ 45 \f{\la t^2\ra^5}{\la t\ra^5}~. \nn
\eea
We now use these in Eq.~(\ref{Qnxn}) to obtain current fluctuations in the ring geometry. 
We immediately see that the moments of ${\cal Q}$ 
 will  now  grow linearly in time. With the cut-off $t_L\sim L^\delta$, we have $\la t^n \ra \sim L^{(2-\beta) \delta}$ and hence we get
\bea
{\la{\cal Q}^2\ra_{ c}  \over t} & \sim &   L^{(2-\beta) \delta-1} \, ,\label{Qnxn-2}\\
{\la{\cal Q}^4\ra_{ c}  \over t} &\sim &  L^{(4-\beta) \delta-3}  \, , \label{Qnxn-4}\\
{\la{\cal Q}^6\ra_{ c}  \over t} & \sim &  L^{(6-\beta) \delta-5}  \, , \label{Qnxn-6}
\eea
and in general we conjecture ${\la{\cal Q}^{2n}\ra_{ c}  / t} \sim   L^{(2n-\beta) \delta-(2n-1)}$.

In one-dimensional mechanical models such as hard-point gas and anharmonic chains, heat is mediated by phonons which are weakly scattered. One can then think of these as performing Levy walks and indeed this picture  
is consistent with  simulation data on energy diffusion \cite{denisov03,metzler04,cipriani05,lepri11}.
Here we now see that the cut-off  time $\tau_L$ 
 also gives  a possible explanation for the behavior seen in simulations on the ring of hard-point alternate gas  of \cite{brunet10}, where the cumulants grow linearly in time with different system size dependence.
There we find that 
$\la{\cal Q}^2\ra_{ c} / t \sim L^{-0.5}$ and $\la{\cal Q}^4\ra_{ c} /t \sim L^{0.5}$. From this one gets 
from Eqs.(\ref{Qnxn-2}) and (\ref{Qnxn-4}) 
\bea
 \beta \sim 5/3 ~~{\rm and}~~ \delta \sim 3/2 \, \nonumber
\eea
which leads  through (\ref{Jsol}) to   a  value
 $\alpha =1/3$ for the anomalous Fourier's law of  the  hard-point alternate gas in the open geometry
consistent with most of the simulations done so far \cite{dhar08,brunet10} for this system.
Furthermore, Eq.(\ref{Qnxn-6}) predicts 
\bea
{\la{\cal Q}^4\ra_{ c}  \over t} \sim   L^{1.5} \, .
\eea
Also we note that the value $\delta= 3/2$ is consistent to the mechanism (c) discussed above which gives $\delta=2/(3-\beta)$, since for this system $\beta=2-\alpha=5/3$.

\section{Discussion}
\label{summary}
The Levy walk model is a natural extension of the ordinary random walk model where we now allow  the walker to move in randomly chosen directions for long time intervals. Several studies suggest that this is could be a good model to describe the motion of phonons in low-dimensional systems and photons in disordered medium. 
In this work we have studied  steady state transport by non-interacting Levy walkers. 
We have computed the average current, the density profile   and the large deviation function of the integrated current, in the open geometry when the system is connected at its two ends to reservoirs.  
Current fluctuations on a ring geometry is also studied where we argue that it is natural to study a Levy walk with a system-size cut-off for the flight time distribution.  
High-dimensional analysis might be relevant for analyzing Levy transport of light through random medium which has been called a Levy glass \cite{barthelemy07}.

{\bf Acknowledgement}: AD  thanks O. Narayan and S. Sabhapandit  for many useful discussions and DST for support through the Swarnajayanti fellowship. KS was supported by MEXT (23740289).

\end{document}